\documentclass[journal=ancac3,manuscript=article]{achemso}
\usepackage[version=3]{mhchem} 
\usepackage{mathptmx,amssymb}
\usepackage{xcolor}
\pdfoutput=1

\author{Vadim I. Zakomirnyi}
\affiliation[Beckman Institute]{Beckman Institute for Advanced Science and Technology, University of Illinois at Urbana-Champaign, Urbana, IL 61801, USA}
\email{vazak@illinois.edu}

\author{Alexander Moroz}
\affiliation{Wave-scattering.com}
\author{Rohit Bhargava}
\affiliation[Depts]{Departments of Bioengineering, Electrical \& Computer Engineering, Mechanical Science \& Engineering, Chemical and Biomolecular Engineering and Chemistry, Cancer Center at Illinois, Beckman Institute for Advanced Science and Technology, University of Illinois at Urbana-Champaign, Urbana, IL 61801, USA}

\author{Ilia L. Rasskazov}
\affiliation[SunDensity]
{SunDensity Inc., Rochester, NY 14604, USA}

\title{Large Fluorescence Enhancement via Lossless All-Dielectric Spherical Mesocavities}

\begin{document}

\begin{tocentry}
\includegraphics[width=7cm]{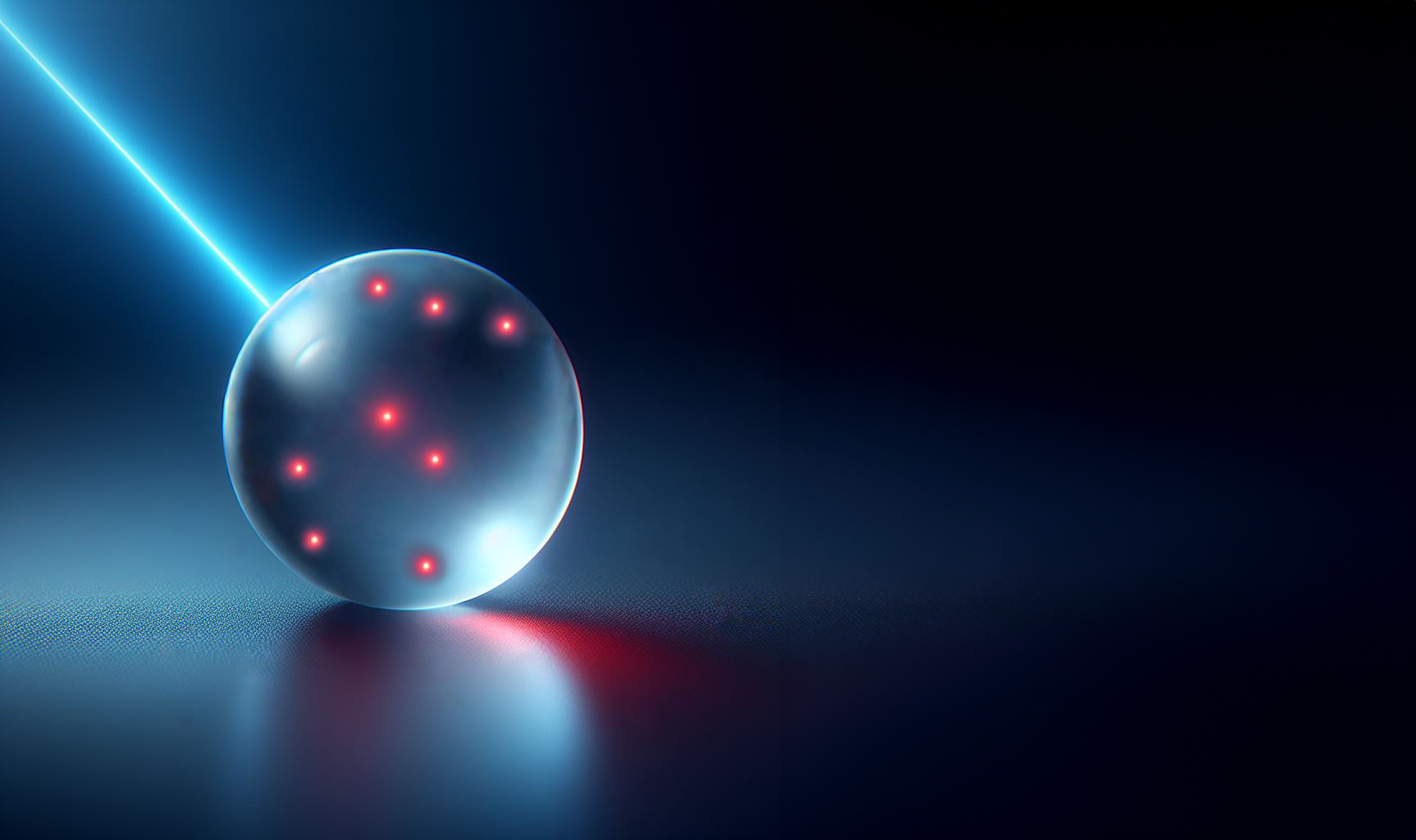}
\end{tocentry}

\begin{abstract}
Nano- and microparticles are popular media to enhance optical signals, including fluorescence from a dye proximal to the particle. Here we show that homogeneous, lossless, all-dielectric spheres with diameters in the mesoscale range, between nano- ($\lesssim 100$~nm) and micro- ($\gtrsim 1$ $\mu$m) scales, can offer surprisingly large fluorescence enhancements, up to $F\sim 10^4$. 
With the absence of nonradiative Ohmic losses inherent to plasmonic particles, we show that $F$ can increase, decrease or even stay the same with increasing intrinsic quantum yield $q_0$, for suppressed, enhanced or intact radiative decay rates of a fluorophore, respectively. 
Further, the fluorophore may be located inside or outside the particle, providing additional flexibility and opportunities to design fit for purpose particles. The presented analysis with simple dielectric spheres should spur further interest in this less-explored scale of particles and experimental investigations to realize their potential for applications in imaging, molecular sensing, light coupling, and quantum information processing.
\end{abstract}


\section{Introduction}
Light-matter interactions are of fundamental importance for advances in a variety of fields, including information technology, sustainable energy, chemical sensing, and spectroscopy. In particular, extensive  research has been driven by the quest to control the electromagnetic effects associated with light at the nanoscale over the past few decades. Plasmonic nanoparticles (NPs) made of noble  (Au, Ag)~\cite{Ostovar2020} and post-transition (Al)~\cite{Knight2014,Gerard2015} metals, transparent conducting oxides (Al:ZnO, Ga:ZnO)~\cite{West2010,Naik2011}, and transition-metal nitrides (TiN, ZrN)~\cite{Naik2011,Naik2012a} are well-known to enhance the electromagnetic field on the nanoscale, enabling extraordinary opportunities for boosting various optical effects and for control over electromagnetic radiation in unusual ways. While immensely successful, the efficiency of optical devices based on plasmonic metal NPs is often  limited due to the Joule, or Ohmic, losses associated with the (free) electron response of metals. Mitigation of these losses has persisted as a challenge and has spurred the search for alternatives. One such route is to use high-quality resonances within all-dielectric particles~\cite{Baranov2017b} for all-dielectric photonics. One of the most exciting and well-developed applications of plasmonic and all-dielectric NPs have been {\em enhanced fluorescence}. Fluorescence underlies many modern detection modalities, especially molecularly-specific recognition of species in low abundance in biological systems. Since intrinsic fluorescence emission can be very weak,  the development of high-performance devices can be challenging for important applications such as single molecule detection~\cite{Stehr2019,Ray2018}, sensitive early diagnostics~\cite{Bower2018,Garcia2018,Park2019a}, bioimaging~\cite{Adachi2023}, detection in complex food and drug safety backgrounds~\cite{Andersen2008}, fingerprint tags~\cite{Lu2018a}, advanced light sources, such as micro/nano light-emitting diodes (LEDs) for high-resolution displays~\cite{Schmidt2017,Yang2015b}, and single photon sources for quantum photonics~\cite{Reimer2019,Xu2019}. Here we address the problem of enhancing fluorescence in order to improve present possibilities and enable potentially promising applications. 

Though large fields around plasmonic NPs are beneficial for fluorescence excitation, unavoidable Ohmic losses reduce the quantum yield. 
As all-dielectric particles do not suffer from these limiting losses, they are a potentially powerful avenue for fluorescence enhancement. However, as widely believed and revealed quantitatively in recent comparative studies~\cite{Sun2016b,Stamatopoulou2021}, all-dielectric particles cannot enhance fluorescence in a manner comparable to plasmonic NPs due to the relatively smaller locally supported electric fields and, consequently, a low enhancement factor. Giant fluorescence enhancement factors reported, thus, are mostly based on a paradigm of \textit{metal}-enhanced fluorescence (MEF), where the use of dielectrics is generally limited to acting as a spacer to keep an emitter at an optimal distance from a metal surface. With almost two decades of intensive development of MEF~\cite{Geddes2002}, the highest recorded fluorescence enhancement factor of $910$ -- $2300$ is considered to be via Ag nanocubes~\cite{Hoang2015,Traverso2021}, which may well be an upper bound of MEF for single particles since Ag is the most favorable plasmonic material for near-field enhancement~\cite{Doiron2019,Sarychev2021}.
Although Mie theory~\cite{Mie1908} has been known for more than a century, rising interest in resonant optical properties of all-dielectric nanostructures has been triggered by recent advances in the fabrication of individual dielectric particles with a controlled geometry~\cite{Baranov2017b}. The resonant behavior of high-index (GaAs, Si, Ge, TiO$_2$) particles enables realization of low-loss nonplasmonic metamaterials and metasurfaces with rich optical functionalities, along with enhanced light–matter interactions and advanced linear and nonlinear light manipulations. Despite an impressive number of obvious advantages of all-dielectric particles, the relatively weaker electric field enhancement compared to plasmonic NPs is widely considered to represent their major drawback, seemingly limiting their utilization. 

Here we demonstrate that all-dielectric particles with sizes in an intermediate region between the nanoscale ($\lesssim 100$~nm) and microscale ($\gtrsim 1$ $\mu$m), or the {\it mesoscale} region, are exceptionally suitable for fluorescence enhancements. The dominant prevalent approaches aim to keep particle sizes as small as possible (e.g. in nanometer range) while accessing higher order resonances. Alas, this leads to conflicting requirements. If the particle size becomes too small, higher order multipole resonances cannot be accessed in the common excitation wavelength ranges for fluorescent species.
When particle sizes become too large, in contrast, whispering gallery mode resonances start to dominate that result in  extremely small line widths~\cite{Lukyanchuk2022,Wang2022meso}. 
Such particles also become too large for many applications due to steric limitations.
By focusing on the mesoscale region, we sought to achieve a compromise between nano- and microworlds while being able to combine the advantages of the both. Our design hypothesis was that mesocavities (MCs) can allow us access to intermediary multipolar resonances ($4 \lesssim l \lesssim  10$) in the visible and near-infrared wavelength ranges. These resonances are well below the typical order of whispering gallery modes ($l \gtrsim 20$) in microparticles~\cite{Lukyanchuk2022,Wang2022meso}; however, their quality factors are much higher than of the orders ($l \lesssim 2$) that can be accessed by nanoparticles.
Such mesoparticles can possess resonances with a quality factor up to $\approx 10^4$ compared to a quality factor of $\approx 20$ for a typical dipole resonance within a plasmonic NP. 
In this manuscript, we use results of our fast numerical open source code Stratify~\cite{Rasskazov20OSAC} 
to design and study mesoscale particles and
to gain insight into their optical responses with a view to optimize their structure and  realize their potential for enhancing fluorescence.

\section{Results and Discussion}

\begin{figure}
    \includegraphics[width=4.3in]{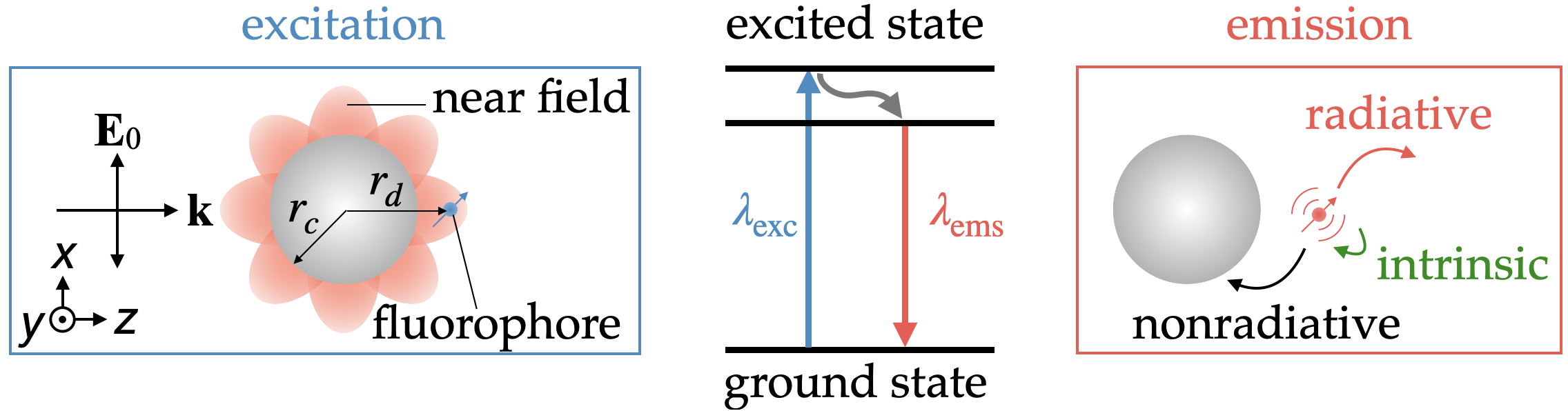}
    \caption{Conventional two-level model for a particle enhanced fluorescence: excitation process under a plane wave illumination and emission process with ED radiation.
    }
    \label{fig:fl_sketch}
\end{figure}

Consider a fluorescence emitter as an oscillating electric dipole (ED)~\cite{Ruppin1982,Ford1984,Chew1987,Chew1988,Kim1988,Moroz2005}. The enhanced fluorescence can be described in two steps as shown in Figure~\ref{fig:fl_sketch}.
In this model, first, a particle locally enhances an electric field, ${\bf E}_{\rm loc}$. This, in turn, amplifies the excitation rate, $\gamma_{\rm exc}$ ($\propto |{\bf E}_{\rm loc}|^2$), of the emitter.
Second, after being excited, the mutual interaction with the particle modifies the radiative decay rate, $\Gamma_{\rm rad}$, of the dipole emitter. The presence of dissipative components also induces changes in the nonradiative decay rate, $\Gamma_{\rm nrad}$. 
Contrary to the widespread misconception that fluorescence enhancement is proportional to radiative decay enhancement, the resulting fluorescence enhancement factor is determined by~\cite{Chew1988,Bharadwaj2007,Sun2016b,Sun2019a}
\begin{equation}
    F = \dfrac{\gamma_{\rm exc}}{\gamma_{\rm exc;0}} \dfrac{q}{q_0} = \dfrac{\gamma_{\rm exc}}{\gamma_{\rm exc;0}}
\dfrac{\tilde{\Gamma}_{\rm rad}}{\tilde{\Gamma}_{\rm rad} + \tilde{\Gamma}_{\rm nrad} + (1-q_0)/q_0 } \dfrac{1}{q_0},
 \label{eq:bff}
\end{equation}
where the subscript ``0" indicates the respective quantity in the free space, the rates with tilde are normalized to the {\em radiative} decay rate in free space (e.g. $\tilde{\Gamma}_{\rm rad}=\Gamma_{\rm rad}/\Gamma_{{\rm rad};0}$, $\tilde{\Gamma}_{\rm nrad}=\Gamma_{\rm nrad}/\Gamma_{{\rm rad};0}$), and $q_0$ is the intrinsic quantum yield. $\tilde{\Gamma}_{\rm rad}$,  which corresponds to the radiative decay enhancement, is determined by the local density of states, whereas $\tilde{\Gamma}_{\rm nrad}$ is determined by particle losses. In the special case of $q_0=1$ and $\tilde\Gamma_{nrad}=0$, the last two fractions on the rhs of eq \ref{eq:bff} reduce to unity. Any fluorescence enhancement is then determined solely by the first fraction, i.e. by the excitation enhancement, and
the fluorescence enhancement in the above particular case is insensitive to any local density of states (LDOS) changes at the emission wavelength. The above illustrates that fluorescence enhancement is a more complex process than radiative decay enhancement and that one should avoid any temptation to characterize fluorescence by either the LDOS or the Purcell factor.

The fluorescence excitation and emission processes are treated independently (i.e., weak coupling) and, in general, there is a Stokes shift between excitation, $\lambda_{\rm exc}$, and emission, $\lambda_{\rm ems}$, wavelengths, i.e. $\Delta\lambda_s=\lambda_{\rm ems}-\lambda_{\rm exc}\neq 0$. The emitter is assumed to be below saturation~\cite{Anger2006,Bharadwaj2007}.
According to eq~\ref{eq:bff}, achieving an optimal fluorescence enhancement factor requires a delicate balance of $\gamma_{\rm exc}$ on one hand, and of $\tilde{\Gamma}_{\rm rad}$ and  $\tilde{\Gamma}_{\rm nrad}$ on the other hand~\cite{Anger2006,Bharadwaj2007,Ringler2008,Arruda2017a,Sun2017b,Sun20JPCC}.

Light-matter interaction of a generic process involving an ED characterized by its ED moment ${\bf d}$ is proportional to $\sim {\bf E}\cdot{\bf d}$. Excitation rates are then  proportional to $|{\bf E}\cdot{\bf d}|^2$. In order to facilitate ensuing calculations, the following common averaging over the ED orientation and its position on a spherical shell of fixed radius $r$ are performed (see Supporting Information for details): 
\begin{itemize}

\item the ED moment ${\bf d}$ in the scalar product $|{\bf E}\cdot{\bf d}|^2$ is averaged over all possible orientations of ${\bf d}$ at a given fixed  ${\bf r}$, whereby $|{\bf E}\cdot{\bf d}|^2$ reduces to $\tfrac13 |{\bf E}|^2 |{\bf d}|^2$

\item after averaging over all possible orientations of ${\bf d}$ at a given fixed  ${\bf r}$, the resulting scalar product  $\tfrac13 |{\bf E}|^2 |{\bf d}|^2$ is averaged over the spherical surface of fixed radius $r$.

\end{itemize}
As a result of this averaging, it is possible to reformulate the scalar product into the respective averages of $|{\bf E}|^2$ and $|{\bf d}|^2$ ~\cite{Bharadwaj2007}. After the above averaging, the excitation rate, $\gamma=|{\bf E}\cdot{\bf d}|^2$, becomes proportional to the surface averaged intensity of the electric field, $\bar \gamma\propto \oint \lvert {\bf E}\rvert^2 \, d{\bf S}$. The latter can be determined by closed-form analytical formulas of ref~\citenum{Rasskazov19JOSAA}. The averaged $\bar \gamma$ is obviously $r$-dependent and does not depend on the spherical angular coordinates. The corresponding decay rates become orientation averaged decay rates~\cite{Moroz2005,Bharadwaj2007} after the above averaging, $\Gamma_{\rm (n)rad} = (\Gamma^\perp_{\rm (n)rad} + 2\Gamma^\parallel_{\rm (n)rad})/3$, where superscripts ``$\perp$'' and ``$\parallel$'' denote the radial and tangential orientations of ED, respectively. All the above steps are implemented in freely available MATLAB code, Stratify~\cite{Rasskazov20OSAC}, employed in the present simulations. In essence, surface integrals of the electric field intensity can be performed analytically and \textit{the calculation of average intensity costs the same computational time as determining intensity at a given single point}.
This strategy allows simulating potentially promising designs at a low computational cost, without the need for time-expensive calculations of electric fields at multiple points inside or outside a particle.
Once configurations of MCs exhibiting large \textit{averaged} values of $|{\bf E}_{\rm loc}|^2$ are identified, further detailed calculations of electric fields for these designs are easily implemented (cf. Figure~\ref{fig:fields}).

The rich modal composition of high-index MCs necessitates a delicate search for suitable resonances. This optimization can be complicated to accomplish as both optimal electric field localization and decay rates need to be optimized for fluorescence enhancement, eq~\ref{eq:bff}. The complexity of this problem is a key challenge why previous works~\cite{Sun2016b,Stamatopoulou2021} did not find the utility of all-dielectric particles for fluorescence enhancement. 

\begin{figure*}[t!]
 \includegraphics{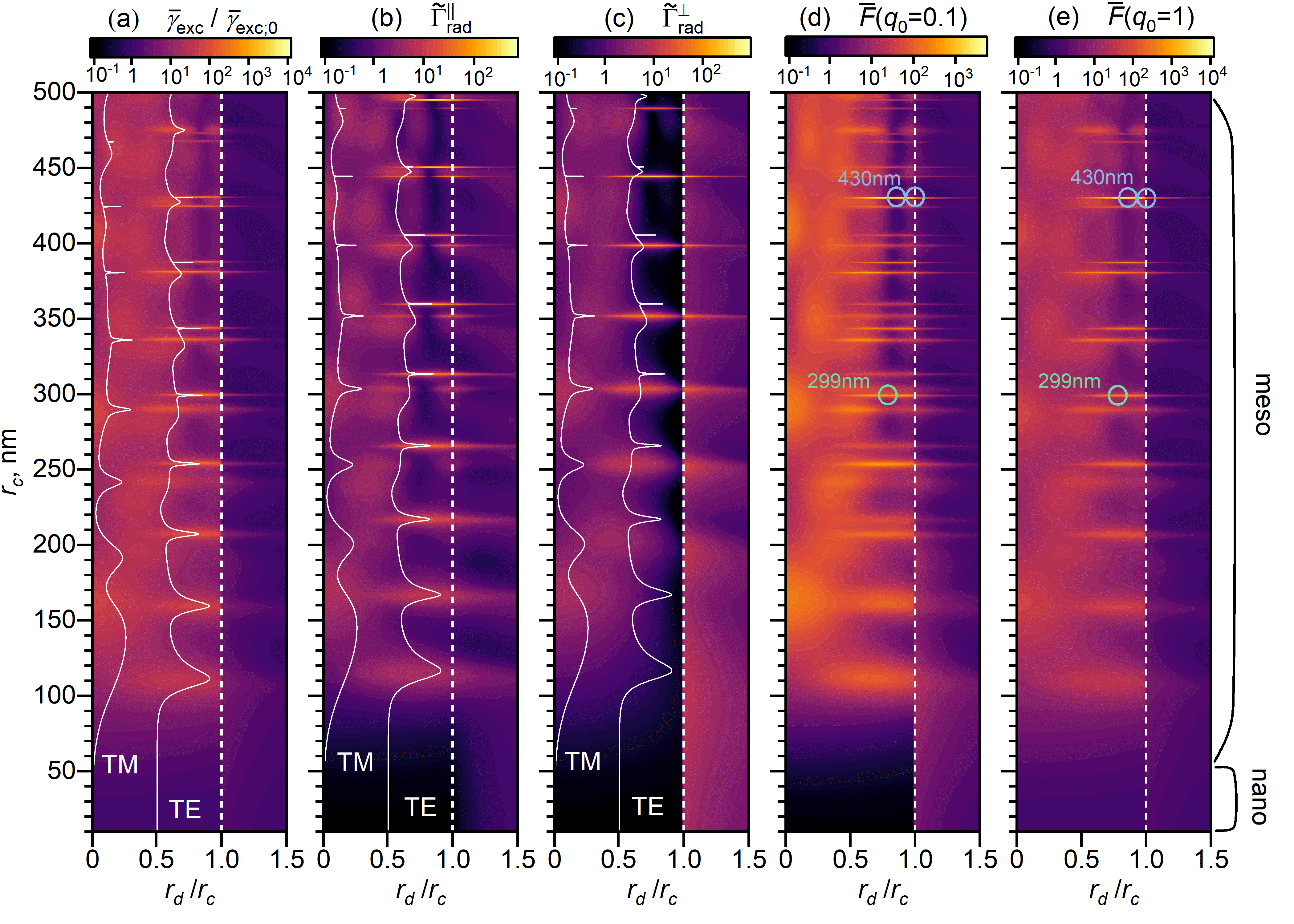}
 \caption{Maps of different quantities as a function of sphere radius, $r_c$, on the $y$-axis and relative ED radial position, $r_d/r_c$, on the $x$-axis for TiO$_2$ MC in the air:
 (a) Averaged excitation rate enhancement. (b)-(c) Normalized
 radiative decay rate for ED with the respective (b) tangential
and (c) radial orientation. (d)-(e)
 Respective averaged fluorescence enhancement $\bar{F}$ for ED with quantum yield (d) $q_0=0.1$ and (e) $q_0=1$.
 Dashed vertical lines at $r_d/r_c=1$ in all plots denote the sphere surface.
 Curves in (a)-(c) show normalized extinction of corresponding TiO$_2$ spheres decomposed into the transverse electric (TE) and transverse magnetic (TM) (with horizontal offset for clarity) modes at $\lambda_{\rm exc}=633$~nm in (a) and at $\lambda_{\rm ems}=663$~nm in (b), (c).
 The peaks in the radiative decay rate correlate nicely with the peaks in the modes having dominant electric field components along the ED direction, i.e. the TE (TM) mode for the tangential (radial) ED orientation.
 Circles in (d) and (e) correspond to configurations considered in detail in Figures~\ref{fig:fields} and \ref{fig:q0}.
 }
 \label{fig:all_maps}
\end{figure*}

\begin{figure}[t!]
 \includegraphics{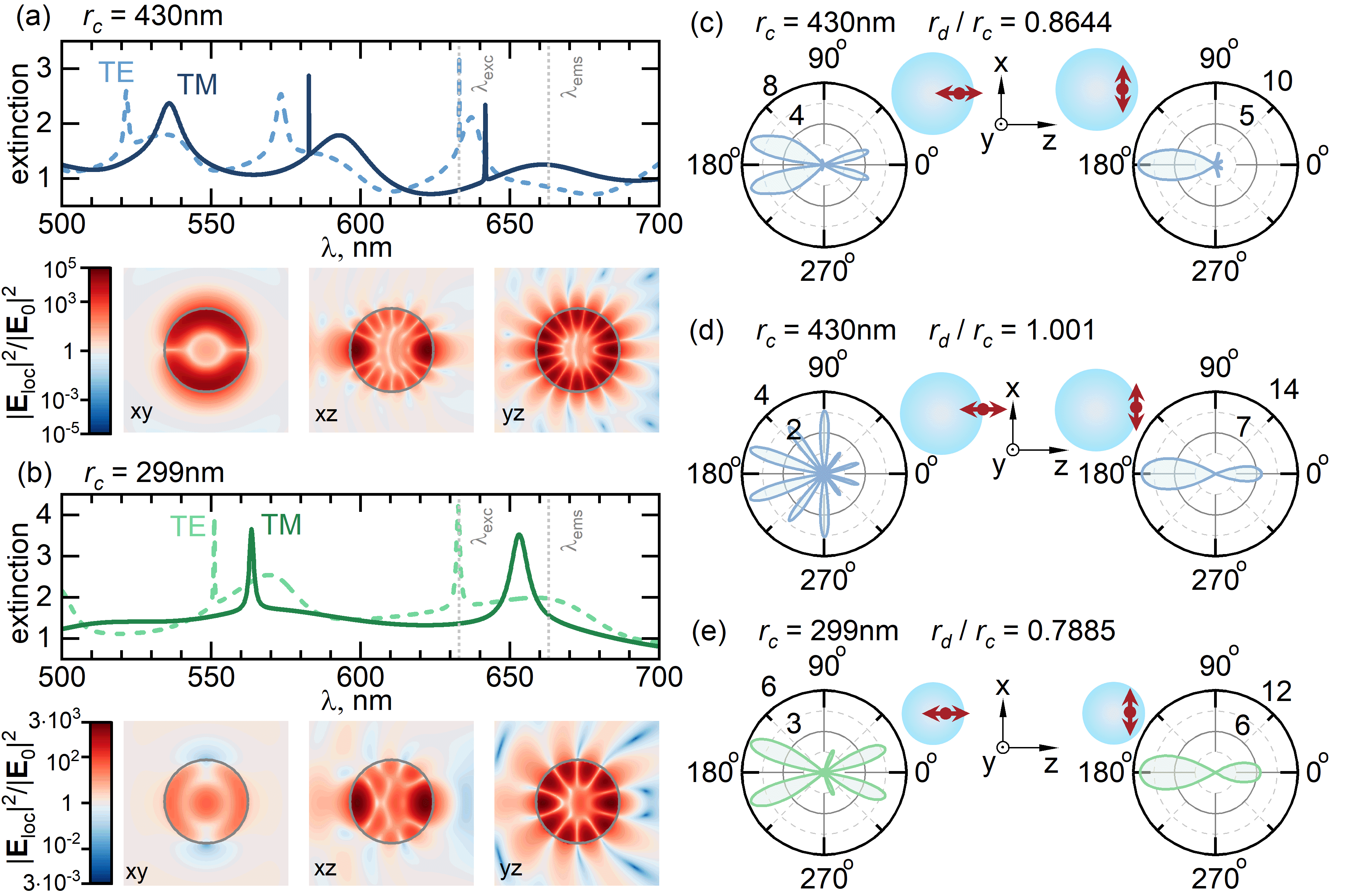}
 \caption{(a-b) Extinction efficiency decomposed into the transverse electric (TE) and transverse magnetic (TM) modes  as a function of wavelength, and normalized intensity of the electric field, $|{\bf E}_{\rm loc}|^2/|{\bf E}_0|^2$, at $\lambda_{\rm exc}$ inside and outside TiO$_2$ sphere with radius (a) $r_c=430$~nm and (b) $r_c=299$~nm. The results correspond to the incident plane-wave amplitude ${\bf E}_0$ oscillating along the $x$ axis and propagation along the $z$ axis as indicated in Figure~\ref{fig:fl_sketch}. 
 Vertical dashed lines in extinction spectra correspond to excitation and emission wavelengths of Nile Blue dye. 
 (c)-(e) Polar plots (in the $xz$ plane) of angle-resolved directivity (in dB) of ED emitting at $\lambda_{\rm ems}=663$~nm for its radial (left -- along the $z$ axis) and tangential (right -- along the $x$ axis) orientations. 
 The following sizes of MC and relative radial positions of the ED are considered (cf. Figure~\ref{fig:all_maps}): (c) $r_c = 430$~nm, $r_d/r_c=0.8644$ (d) $r_c = 430$~nm,  $r_d/r_c=1.001$ and (e) $r_c = 299$~nm, $r_d/r_c=0.7885$. The directivity varies with the angle $\theta$, being oriented to the free space for $\theta=0^\circ$ and toward the center of the particle for $\theta=180^\circ$.}
 \label{fig:fields}
\end{figure}

The {\em averaged} fluorescence enhancement factor (eq~\ref{eq:bff}) in the presence of a lossless ($\tilde{\Gamma}_{\rm nrad}=0$) all-dielectric particle reduces to
\begin{equation}
 \bar F = \dfrac{\bar\gamma_{\rm exc}}{\bar\gamma_{\rm exc;0}} \dfrac{\tilde{\Gamma}_{\rm rad}}{1-q_0 + q_0 \tilde{\Gamma}_{\rm rad}}\cdot
 \label{eq:fl_lossless}
\end{equation}
To relate to practical applications, we consider widely used and readily fabricated TiO${}_2$ spherical particles with real refractive index $n_c=2.7$, which are lossless in the visible range~\cite{Baranov2017b}, and vary their radii from $10$ to $500$~nm, thus sampling both the ``nano'' (radii from $10$ to $50$~nm) and ``meso'' regimes (radii from $50$ to $500$~nm). 
For the sake of illustration, either 
standard Nile Blue dye with $\lambda_{\rm exc}=633$~nm and $\lambda_{\rm ems}=663$~nm, or Er${}^{3+}$ with $\lambda_{\rm exc}=488$~nm and $\lambda_{\rm ems}=1530$~nm is used as ED emitter. It is also worth noting that modern fabrication techniques can embed emitters into spheres with nanometer precision in the radial direction for a number of different materials~\cite{VanBlaaderen1992,Gritsch2022}. Controlled positioning of emitters outside the sphere is possible by attaching emitters to (nano)particles via molecular techniques, such as using single-stranded DNA (ssDNA) spacers~\cite{Dulkeith2005} or DNA origami~\cite{Acuna2012}. Therefore, we consider situations with an emitter both inside and outside of a TiO$_2$ sphere.
Figures~\ref{fig:all_maps}(a-c) show excitation and radiative decay rates for two different ED orientations. Extinction spectra for the 
transverse electric (TE), also called magnetic, and transverse magnetic (TM), also called electric, modes are superimposed on the top of the graphs to show the nature of resonances. The peaks in the radiative decay rate correlate nicely with the peaks in the modes having dominant electric field components along the ED direction, i.e. the TE (TM) mode for the tangential (radial) ED orientation.

Similar to our previous study~\cite{Rasskazov21JPCL}, we observe that it is more beneficial to tune a suitable resonance to an emitter excitation wavelength than to its emission wavelength for enhancing $\bar F$. This is because an enhancement in $\tilde\Gamma_{\rm rad}$ does not need to have any effect on enhancing fluorescence. 
In view of eq~\ref{eq:fl_lossless}, $\bar F$ is proportional to $\tilde\Gamma_{\rm rad}$ only for intrinsic quantum yield $q_0\equiv 0$, whereas $\bar F$ is {\em independent} of $\tilde\Gamma_{\rm rad}$ for $q_0=1$~\cite{Rasskazov22Plas}. 

Tuning a suitable {\em magnetic} (TE) MC resonance to the excitation wavelength was found to be the most beneficial. The so-called decay rate engineering advocated within the MEF framework as means of enhancing $F$~\cite{Lakowicz2002,Geddes2002} plays a smaller role in the present case of MCs. 

In the interstitial case described by the virtual-cavity model~\cite{Vries1998}, the local field, ${\bf E}_{\rm loc}$, felt by emitters \textit{inside} the particle in the presence of a macroscopic field ${\bf E}$ is ${\bf E}_{\rm loc}=L_{\rm vc} {\bf E}$, where
\begin{equation}
L_{\rm vc}=\dfrac{n_c^2+2}{3}
\end{equation}
is the so-called Lorentz local-field correction~\cite{Vries1998} (see Supporting Information for its general derivation from the Maxwell's equations).
This correction is particularly interesting for high-index dielectrics because $L_{\rm vc}$ linearly increases with the host dielectric constant, and it can become large. Averaged fluorescence enhancements $\bar{F}$ displayed in Figures~\ref{fig:all_maps}(d),(e) for ED  for two different values of $q_0$ ($0.1$ and $1$) are shown with the Lorentz local-field correction included (contributing for $n_c=2.7$ the factor of $L_{\rm vc}^2\approx 9.6$ to $\bar\gamma_{\rm exc}$).

Figure~\ref{fig:fields}a,b shows extinction efficiency and normalized intensity of the electric field, $|{\bf E}_{\rm loc}|^2/|{\bf E}_0|^2$ (at $\lambda_{\rm exc}$), inside and outside TiO$_2$ sphere with radius (a) $r_c=430$~nm and (b) $r_c=299$~nm, on the background of the respective TE and TM resonances. 
The results correspond to the incident plane-wave amplitude ${\bf E}_0$ oscillating along the $x$ axis and propagation along the $z$ axis, as indicated in Figure~\ref{fig:fl_sketch}. 

Local fields enhancement demonstrates rarely studied patterns of $l=8$ and $l=5$ multipoles induced in MCs with $r_c=430$~nm and $r_c=299$~nm, respectively.
We notice that optimal arrangement of emitters into hot spots of a mesosphere can further increase the ultimate fluorescence enhancement factor.
Interestingly, the spatial extent of these hot spots is quite broad, which makes them suitable for any application benefiting from strong electric field enhancement.
 
Coupling of ED to a dielectric mesoparticle can result in strong modification of angular emission of ED. To analyze it, we have calculated the radiation directivity (relation of the power emitted into a certain direction to the solid angle averaged emitted power) of ED in the vicinity of TiO${}_2$ mesoparticle at $\lambda_{\rm ems}=663$~nm, which corresponds to the maximum of emission wavelength of Nile Blue dye. 
Figure~\ref{fig:fields}(c)-(e) shows the angular directivity of radially ($\perp$) and tangentially ($\parallel$) oriented ED at specific relative positions highlighted in Figure~\ref{fig:all_maps}.
It can be seen that radially oriented ED doesn't have large directionality.
Tangentially oriented ED, on the other hand, has a highly directional pattern of emission oriented toward mesoparticle ($\theta=180^\circ$).
Interestingly, $\approx 12$ directivity for ED with tangential orientation embedded inside the mesoparticle with $299$~nm (Figure~\ref{fig:fields}(e)) approaches the fundamental Harrington-Chu limit~\cite{Harrington1958,Chu1948} ($\approx 13.69$ for a given sphere size and $\lambda_{\rm ems}$).

\begin{figure}
 \includegraphics{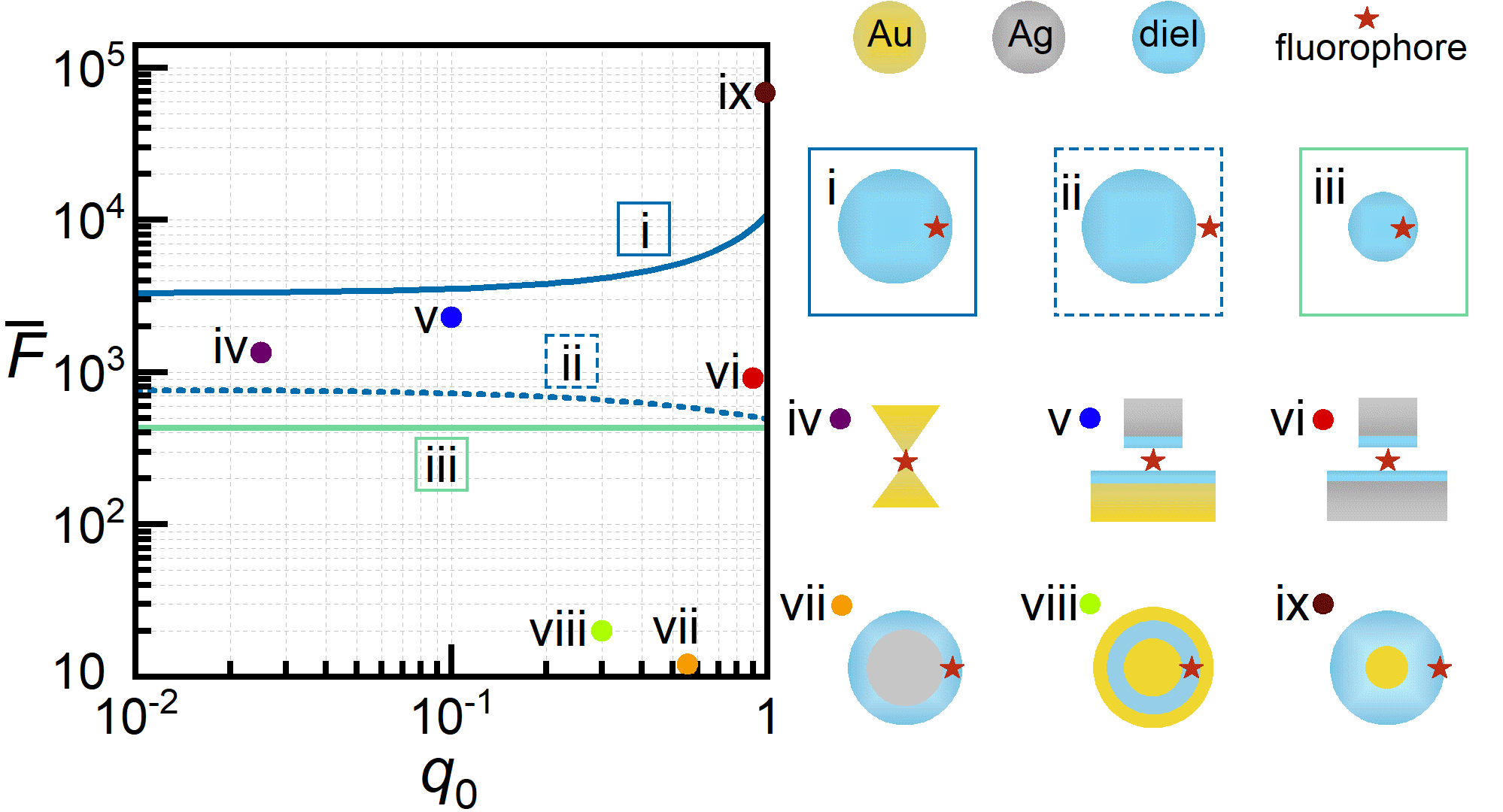}
 \caption{Averaged fluorescence enhancement as a function of $q_0$ for three different TiO$_2$@Nile Blue configurations highlighted in Figure~\ref{fig:all_maps}d,e:
 (i) $r_c=430$~nm and $r_d/r_c=0.8644$,
 (ii) $r_c=430$~nm and $r_d/r_c=1.001$,
 (iii) $r_c=299$~nm and $r_d/r_c=0.7885$.
 $\tilde{\Gamma}_{\rm nrad}\equiv 0$ implies the absence of familiar $\bar F \sim 1/q_0$ scaling within the MEF framework~\cite{Rasskazov22Plas}.
 For comparison, the most representative cases of fluorescence enhancements and respective $q_0$ described in previous studies are shown:
 (iv) bow-tie Au antenna~\cite{Kinkhabwala2009},
 (v) Ag nanocubes on Au film~\cite{Hoang2015},
 (vi) Ag nanocubes on Ag film~\cite{Traverso2021},
 (vii) Ag@dielectric core-shells~\cite{Tovmachenko2006},
 (viii) Au@dielectric@Au matryoshkas~\cite{Ayala-Orozco2014a},
 (ix) Au@dielectric mesoscale core-shell~\cite{Rasskazov21JPCL} (the result therein has been multiplied by the factor of $22.56$, which corresponds to $L_{\rm vc}^2$ for the dielectric shell with $n_s=3.5$ considered in ref~\citenum{Rasskazov21JPCL}).
 The comparison with other geometries is to illustrate the potential of our simple geometry at a different scale. In addition to the optimized geometries for each shape, an end user must also consider the constraints of the application to choose one of these options.
 }
 \label{fig:q0}
\end{figure}

 If in the lossless case described by eq~\ref{eq:fl_lossless}: 
(i) $\tilde\Gamma_{\rm rad}>1$ then $q/q_0$ (the second fraction in eq~\ref{eq:fl_lossless}) grows as $q_0$ decreases; 
(ii) $\tilde\Gamma_{\rm rad}<1$ then $q/q_0$ grows as $q_0$ increases; 
(iii) $\tilde\Gamma_{\rm rad}=1$ then $q/q_0$ does not depend on $q_0$. As shown in Figure~\ref{fig:q0}, one can encounter all three cases in practice.
Contrary to the case of plasmonic particles characterized by significant $\tilde{\Gamma}_{\rm nrad}$ in their proximity due to large Ohmic losses leading to a scaling $\bar F \sim 1/q_0$ behavior in $q_0$ \cite{Rasskazov22Plas}, we have $\tilde{\Gamma}_{\rm nrad}=0$  here and, consequently, there is no scaling in $q_0$. 
In fact, the maximum achievable $\bar F$ can be increasing (i.e. not decreasing) with increasing $q_0$ as can be seen from Figure~\ref{fig:q0}.

Enhancement mediated by dielectric nanoparticles is a common approach to increasing the intensity of fluorescence signals emitted by fluorescent molecules or materials. When fluorescent molecules are placed near dielectric nanoparticles or embedded in them, the nanoparticles can act as efficient scattering elements, increasing the excitation and$/$or emission rate of these fluorescent molecules through the enhanced local field effect. 
The choice between dielectric microparticles and nanoparticles for the purpose of fluorescence enhancement depends on the specific requirements of the application, such as the desired enhancement factor, the photostability of the fluorescent molecule, and compatibility with the experimental conditions.
In our study, we show that much larger fluorescence enhancement can be achieved with meso-sized particles than with nanoparticles. Contrary to earlier studies involving all-dielectric microparticles~\cite{Lukyanchuk2022,Wang2022meso}, our study has been strictly limited to particle sizes between nanoscale ($\lesssim 100$~nm) and microscale ($\gtrsim 1$ $\mu$m) and, consequently, to intermediary multipolar resonances ($4 \lesssim l \lesssim  10$) that is well below typical order of whispering gallery modes ($l \gtrsim 20$) in microparticles. The present proposal seems to be an excellent compromise between nano- and microworlds while being able to combine the advantages of both worlds.

Recently, we have reported that a metal core is essential to get extraordinary fluorescence enhancements exceeding $3000$ (results presented therein are shown without taking into account the local field corrections)~\cite{Rasskazov21JPCL}. 
The reported fluorescence enhancement by homogeneous all-dielectric MCs is only slightly smaller than that
reported in ref~\citenum{Rasskazov21JPCL} (cf. Figure~\ref{fig:q0}), while bringing significant manufacturing advantages: (i) homogeneous all-dielectric MC are generally much easier to fabricate, (ii) do not require costly noble metals, and  (iii) allow one to embed emitters in their entire volume (or realize other options~\cite{VanBlaaderen1992}) potentially leading to brighter fluorescence sources. Dielectric MCs can provide a tunable and uniform enhancement effect over a wide range of wavelengths. This is because the resonant frequency of the dielectric MCs can be tailored to match the excitation and emission wavelengths of the fluorescent molecule, leading to a more efficient energy transfer.
\begin{figure}[hbt!]
 \includegraphics[width=3.33in]{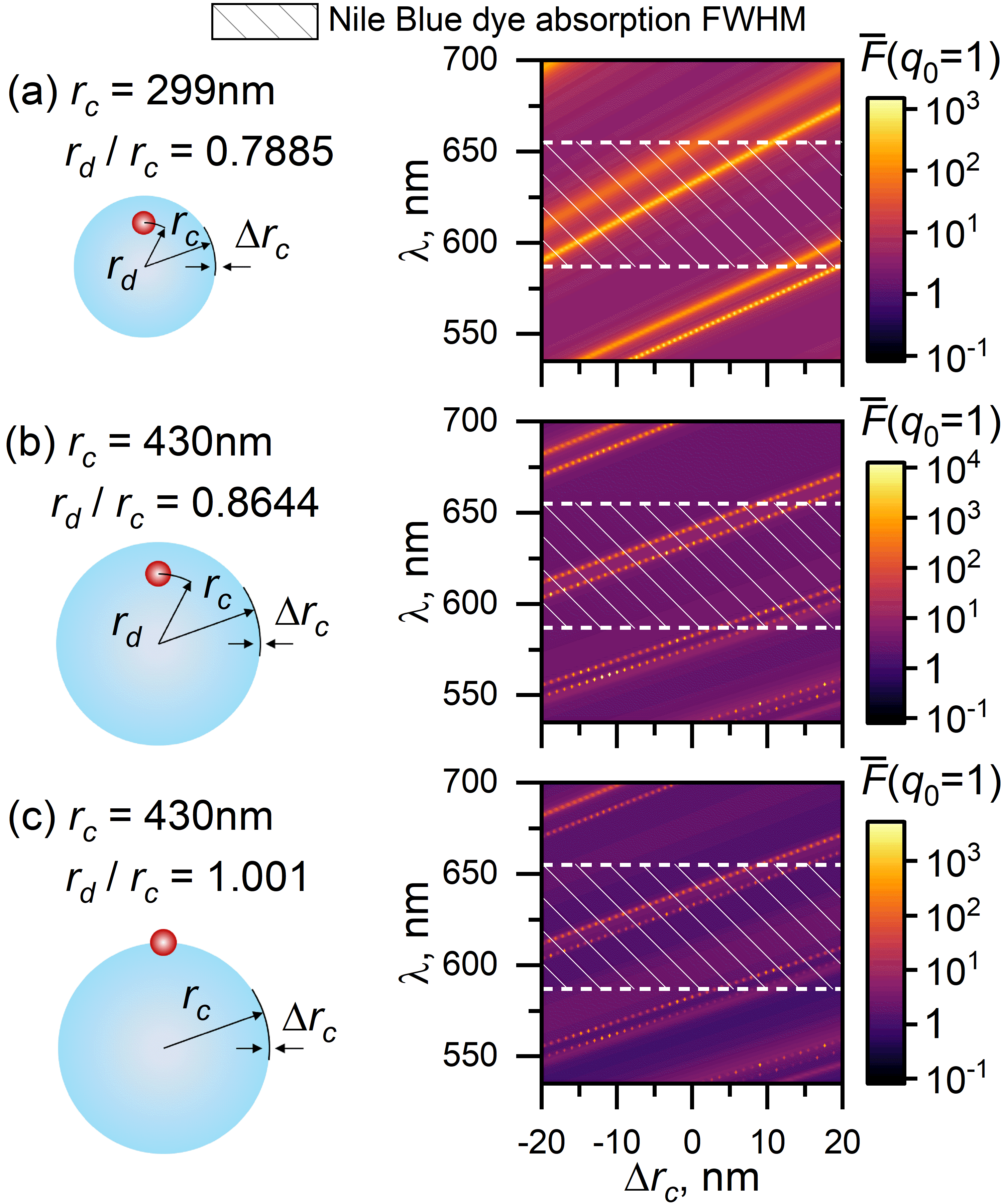}
 \caption{
 Averaged fluorescence enhancement $\bar{F}$ as a function of wavelength and of the deviation, $\Delta r_c$, in the size of MC from optimally chosen (cf. Figure~\ref{fig:all_maps}): 
 (a)~$r_c=299$~nm,
 (b)~$r_c=430$~nm, and
 (c)~$r_c=430$~nm.
 As $\Delta r_c$ varies excitation is assumed to be tuned to the corresponding resonance by a tunable (e.g. laser diode) source.
 Corresponding relative positions of ED are fixed to be the same for different $\Delta r_c$ values and are denoted in the respective sketches.
 The shaded area between horizontal dashed lines in all plots denotes the fwhm of Nile Blue dye absorption spectra. 
  }
 \label{fig:size_dispersion}
\end{figure}

We next discuss practical aspects of utilizing mesoparticles for fluorescence enhancement (Figure~\ref{fig:size_dispersion}).
We start with examining the effect of particle size polydispersity on the practically achievable fluorescence enhancements.
Given that a typical full width at half maximum (fwhm) of organic dyes absorption band is at least $50$ nm~\cite{Drexhage1973,Drexhage1974}, and assuming as large as $\pm 5$ nm range of particle size distribution about a target size, the latter size variations can be easily accommodated by a tunable laser diode source, enabling one to excite dyes on the particle resonance, while still remaining within a dye absorption band. 
Figure~\ref{fig:size_dispersion} shows that the effect of the particle size polydispersity under the above provisions is rather small.
As a matter of fact, even as large as $\pm 20$ nm range of particle size distribution about a target size would be sufficient for enjoying enhanced fluorescence for most applications. We have also observed another interesting result from this figure: high fluorescence enhancement values can be achieved for all possible values of $\Delta r_c$ within the range of predefined $r_c$ and $r_d/r_c$. This is enabled by the high multipole order of resonance of the particles, occurring at a wavelength that coincides with the maximum of fluorescence. It is clear that the shaded area between the horizontal dashed lines, which corresponds to the fwhm of the absorption spectrum of Nile Blue dye, will always be enhanced by the presence of $5$th-$8$th order TE or TM resonances of dielectric MC.

\begin{figure}[hbt!]
 \includegraphics{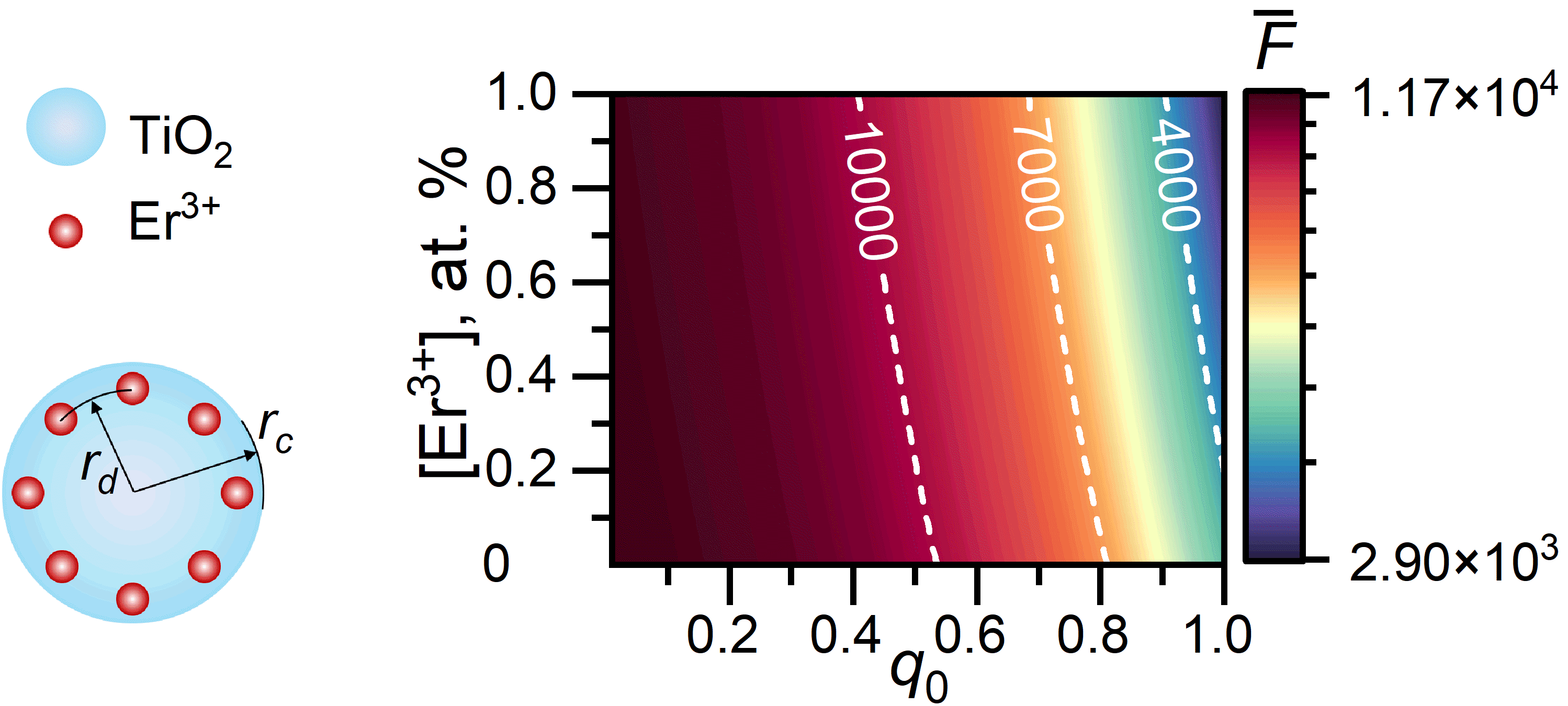}
 \caption{Averaged fluorescence enhancement $\bar{F}$ for TiO$_2$@Er$^{3+}$ composites as a function of the ${\rm Er}^{3+}$ concentration and intrinsic quantum yield $q_0$.
 TiO$_2$ sphere of radius $r_c=489$~nm embedded in water ($n_h=1.33$) with ${\rm Er}^{3+}$ emitters located at fixed $r_d=438$~nm distance from the sphere center.}
 \label{fig:Er}
\end{figure}
While we have considered a single emitter thus far, other emitters and species may also be proximally present. Let us consider the effect of multiple emitters (Figure~\ref{fig:Er}). Once an emitter is excited, it can transfer its excitation to a nearby emitter via an energy transfer mechanism. With multiple emitters present, this transfer process increases the probability of a decay of emitter excitation. Similarly, if the excitation is transferred to an emitter in a proximity of an impurity, the excitation can be quenched (e.g., OH bonds in silica), whereby the excitation disappears without ever contributing to radiation. 
This is the essence of the so-called concentration quenching, which does contribute to nonradiative decay rate but whose origin is different from the Ohmic losses of metal particles.
In what follows, we consider Er${}^{3+}$ emitters with $\lambda_{\rm exc}=488$~nm and $\lambda_{\rm ems}=1530$~nm coupled to TiO$_2$ mesoparticle embedded in water with $n_h=1.33$.
Concentration quenching in this case can be accounted by~\cite{DeDood2001}
\begin{equation}
{\Gamma}_{\rm nrad} = 8\pi C_{\rm Er-Er}[{\rm Er}^{3+}][Q],
\end{equation}
where [Er$^{3+}$] ([Q]) is the erbium (quencher) concentration in at. \%, and $C_{\text{Er–Er}}$ is a coupling constant.
Using a typical coupling value for $C_{\text{Er–Er}}$ from the
literature ($10^{-39}$ cm${}^6$ s${}^{-1}$), we also assume a quencher concentration of 100 ppm for representative calculations. 
For example, assuming the dominant quencher in SiO${}_2$ is OH, this is a reasonable value for the colloids prepared in a wet–chemical reaction~\cite{DeDood2001}.
The effect of Er$^{3+}$ concentration on the averaged fluorescence enhancement $\bar{F}$ is highlighted in Figure~\ref{fig:Er}. Interestingly, even relatively small nonradiative losses have the effect similar to that in the MEF~\cite{Rasskazov22Plas}, namely larger $\bar{F}$ are attainable with smaller $q_0$ .
In the case of non-zero nonradiative losses, utilization of the emitters with the Stokes shift, $\Delta \lambda_s$, smaller than the resonance line width can be beneficial allowing one to achieve large $\bar\gamma_{\rm exc}$ and $\tilde{\Gamma}_{\rm rad}$ simultaneously, thus counteracting any emergent quenching induced by $\tilde{\Gamma}_{\rm nrad} \neq 0$.

\section {Conclusions}
By using our fast calculation approach, modeling showed that {\em size matters}: for a given sphere of specific refractive index, much larger fluorescence enhancement can be achieved with meso-sized homogeneous dielectric spheres than with nanospheres. Quite surprisingly, record high averaged fluorescence enhancements for all-dielectric particles of $\bar{F}\sim 10^4$ can be achieved with simple homogeneous mesospheres under optimal conditions without  requiring the engineering of sophisticated shapes, precise nanogaps, generation of hot spots, or designer metasurfaces. Potential fluorescence enhancements are much larger than those obtained experimentally (Figure~\ref{fig:q0}). Given the importance of fluorescence in numerous practical applications, we find it a very important result beneficial for fluorescence applications to imaging, sensing, strong coupling, and quantum information processing. Only averaged quantities have been discussed so far; it is to be expected that fluorescence enhancements of individual fluorophores can be a further magnitude larger if they are optimally arranged into hot spots of a mesosphere.

\section {Methods}

Numerical results reported in this manuscript have been obtained via open source code Stratify~\cite{Rasskazov20OSAC,stratify}.
Detailed derivations and the respective expressions for quantities entering master eq~\eqref{eq:fl_lossless} are available as follows: $\bar{\gamma}_{\rm exc}/\bar{\gamma}_{\rm exc;0}$~\cite{Bott1987,Rasskazov19JOSAA} and $\tilde{\Gamma}_{\rm rad}$~\cite{Ruppin1982,Chew1987,Moroz2005}.

\begin{suppinfo}
Averaged fluorescence enhancement as a proper figure of merit.
A general derivation of the Lorentz local-field correction directly from the Maxwell's equations. 
\end{suppinfo}

\textbf{Associated Content}: Vadim I. Zakomirnyi; Alexander Moroz; Rohit Bhargava; Ilia L. Rasskazov. Large fluorescence enhancement via lossless all-dielectric spherical mesocavities. 2023. arXiv. https://arxiv.org/abs/2301.10899 (accessed December 1, 2023)

\providecommand{\latin}[1]{#1}
\makeatletter
\providecommand{\doi}
  {\begingroup\let\do\@makeother\dospecials
  \catcode`\{=1 \catcode`\}=2 \doi@aux}
\providecommand{\doi@aux}[1]{\endgroup\texttt{#1}}
\makeatother
\providecommand*\mcitethebibliography{\thebibliography}
\csname @ifundefined\endcsname{endmcitethebibliography}
  {\let\endmcitethebibliography\endthebibliography}{}

\end{document}


\usetagform{supplementary}

\section{Averaged fluorescence enhancement}
The excitation rate $\gamma_{exc}$ is governed by the electric dipole interaction term in the Hamiltonian $\sim {\bf E}\cdot{\bf d}$. The rate itself is proportional to $\sim |{\bf E}\cdot{\bf d}|^2$. At any given radial position, the averaging over all possible dipole orientations, while maintaining dipole magnitude, yields $\sim E_i E_j \overline{d_i d_j}$, where the summation over repeated indices is assumed. For unequal indices, $i\ne j$, to any $d_id_j$ there corresponds another $d_id_j$ with equal magnitude but opposite sign.
Therefore upon averaging over all possible dipole orientations at a given radial position $E_i E_j \overline{d_i d_j}$ reduces to $E_j^2 \overline{d_j^2}$. But the averaged components $\overline{d_j^2}$ are all the same and equal to $\tfrac13 |{\bf d}|^2$.
Thus at the end $E_j^2 \overline{d_j^2}$ reduces to $\tfrac13 |{\bf E}|^2 |{\bf d}|^2$. Ensuing averaging over all possible radial orientations, which is averaging over spherical shell of fixed radius $r$, affects only $|{\bf E}|^2$ but not $|{\bf d}|^2$. This ensures the decoupling of ${\bf E}$ and ${\bf d}$ in the averaging. When forming normalized dimensionless quantities, i.e. $\tilde\gamma_{exc}$ as the ratio of $\gamma_{exc}$ with and without a sphere, the above factor $\tfrac13 |{\bf d}|^2$ cancels out, because it is the very same in the respective cases. Consequently, $\tilde\gamma_{exc}$ is entirely determined by the enhancement of the averaged electric intensity.

In a typical device (e.g. fluorescence probes for confocal imaging, tagging and tracing of organic molecules, etc) there is always a finite distribution of fluorescence emitters. Dipole moments of the emitters are as a rule also distributed randomly (e.g. ion beam implantation of rare earth emitters). Therefore, as far as devices are concerned, there are the averaged quantities which are relevant for their description. The averaging provides a relevant figure of merit in order to assess the fluorescence of a system under consideration. It is obvious that higher fluorescence enhancement can be achieved in practice by exercising a control over fluorophore dipole orientation and by judicious positioning of a fluorophore in suitable hot spots.

\section{Local field corrections}
The local field within the cavity, ${\bf E}_{loc}$, differs from an applied macroscopic field, ${\bf E}$ by a local-field correction factor $L$, ${\bf E}_{loc}=L {\bf E}$.
Local field corrections inside dielectrics exhibit the well-known (i) {\em real}, or {\em empty-cavity}, and (ii) {\em virtual-cavity}, or {\em Lorentz local-field}, factors for {\em substitutional} and {\em interstitial} atoms, respectively~\cite{Vries1998}. 

The real cavity model assumes that (i) the atom is at the 
center of the cavity and (ii) the cavity itself has no other material, i.e. it is empty. The resulting ratio of local and macroscopic fields is
given by [cf. electrostatic result \rfs{elocinc} below]
\begin{equation}
1 <  L_{rc}= \frac{3\varepsilon}{2\varepsilon+1} < \frac32 \cdot
\label{Lrc}
\end{equation}
The substitutional case occurs prevalently for impurity atoms \cite{Vries1998} and rare-earth emitters embedded within different organic complexes~\cite{Duan2005}, whenever the emitter embedded in a dielectric host expels the dielectric media and creates there a real tiny cavity.
As pointed out by B\"ottcher based on the initial work of Onsager in 1933,
the local-field correction factor depends also on the polarizability $\chi$ of the molecule placed in the cavity. The real cavity model applies
when the polarizability of the guest dipole is sufficiently low compared to that of the host so that the reaction field caused by the induced dipole acting on the cavity can be neglected~\cite{Aubret2019}. 

The {\em virtual} cavity model assumes a uniform distribution of material within and outside the cavity~\cite{Aspnes1982,Vries1998,Dolgaleva2012,Aubret2019}. 
The corresponding local-field correction is then known as the Lorentz local-field correction, eq~3, 
\begin{equation}
1< L_{vc}= \frac{\varepsilon+2}{3}\cdot
\label{Lvcs}
\end{equation}

For pure systems constituted of only one kind
of atom, or when the polarizability of the guest is the same as that of the host, the interstitial case of Ref.~\cite{Vries1998} described by the virtual-cavity model applies. This is also the case of rare earth emitters implanted in dielectrics by ion beam deposition~\cite{DeDood2001a}.

One can notice a significant difference between the respective local-field corrections \rfs{Lrc} and eq~3: whereas $L_{rc}$ is strictly bounded by $3/2$ from above, the Lorentz local-field correction \rfs{Lvcs} increases indefinitely with increasing $\varepsilon$ and is, in principle, unbounded from above. In general, $L_{rc}$ represents a lower bound for the local field factor~\cite{Aubret2019}.

The textbook derivation of local-field corrections is usually performed within quasi-static approximation~\cite{Aspnes1982,Vries1998,Dolgaleva2012,Aubret2019}. Let us consider a sphere with dielectric constant $\varepsilon_s$ embedded in a host characterized by dielectric constant $\varepsilon_h$. Irrespective of the units used, one finds for the sphere in the electrostatic case
\begin{equation}
{\bf E}_{loc} = \frac{3\varepsilon_h}{\varepsilon_s + 2\varepsilon_h}\, {\bf E}_{inc}.
\label{elocinc}
\end{equation}
The above familiar electrostatic result illustrates the necessity of applying local field corrections when estimating the local excitation field ${\bf E}_{loc}$ felt inside a {\em real} cavity by atoms and molecules in the presence of a macroscopic field ${\bf E}_{inc}$.

Less known is a general derivation of the Lorentz local-field correction directly from the Maxwell's equations.
In general, i.e. beyond the electrostatic approximation, 
the origin of the local-field factors is that the conventional electric dyadic Green's function is {\em not} sufficient to determine the correct value of ${\bf E}({\bf r})$ at source points~\cite{Yaghjian1980}. One has
\begin{equation}
{\bf E}({\bf r}) =i\omega\mu_0\, \lim_{\delta\rightarrow 0} \int_{V_j-V_\delta} {\bf G}({\bf r},{\bf r}')\cdot{\bf J} ({\bf r}')\, dV'
+ \frac{{\bf L}\cdot{\bf J}({\bf r})}{i\omega\varepsilon_0 },
\label{elfvl}
\end{equation}
where $V_\delta$ is an excluded volume, i.e. a cavity comprising the observation point {\bf r}, ${\bf G}$ is the Green's function to the equation
\begin{equation}
[\mbox{\boldmath $\nabla$}\times\mbox{\boldmath $\nabla$}\times - k^2] {\bf G} = \delta({\bf r}-{\bf r}'){\bf I},
\label{vgdef}
\end{equation}
and ${\bf L}$ is an extra dyadics~\cite{Yaghjian1980}
\begin{equation}
{\bf L}=\frac{1}{4\pi}\oint_{S_\delta} \frac{{\bf n}\otimes {\bf E}_{R'}}{R'^2}\, dS'.
\label{vlform}
\end{equation}
Here ${\bf n}$ is the unit normal pointing out of the principal volume 
and ${\bf E}_{R'}$ is the unit vector pointing from ${\bf r}$ to ${\bf r}'$.
The integral on the rhs of \rfs{elfvl} is an {\em improper} nonconvergent integral in the sense of Kellogg~\cite{Kellogg1967}, i.e. it is necessary to restrict the shape of $V_\delta$ in order to obtain a limit when $V_\delta\to 0$. 
Nevertheless, although each of the two contributions on the rhs of \rfs{elfvl} does individually depend on the shape of of excluded volume $V_\delta$, the rhs of \rfs{elfvl} as a whole is independent of the shape of $V_\delta$.
For arbitrary principal volumes and time harmonic fields,
${\bf L}$ can be concisely interpreted physically as a generalized depolarizing dyadics yielding the ``local field" ${\bf E}_0 + {\bf L}\cdot{\bf J}/\varepsilon_0$ of electrostatics. Conversely, the  depolarizing factors for an ellipsoid and the generalization to arbitrary shaped holes in or bodies of uniform volume sources can be found by the formula \rfs{vlform}.

Suppose one were to measure the electric field at a point within 
an enforced current distribution by removing an infinitesimally
small volume $V_\delta$ of current and inserting an ideal point probe therein.
Then the ${\bf L}$ term determines the perturbation in electric field
caused by the hypothetical measurement scheme of removing 
an infinitesimally small volume $V_\delta$ of enforced uniform current. 
Indeed, the measured field would then be given by \rfs{elfvl} but without the 
${\bf L}$ term, since the enforced current at this point has been removed.
The measured local field would then be~\cite{Yaghjian1980}
\begin{equation}
{\bf E}({\bf r}) - \frac{{\bf L}\cdot{\bf J}}{i\omega\varepsilon_0 },
\label{elfvloc}
\end{equation}
and would depend upon the shape of the infinitesimal volume and its relative 
position and orientation with respect to the point probe.
Provided that ${\bf J}$ here is merely an enforced polarization current, and assuming harmonic $e^{-i\omega t}$ time dependence,
\begin{equation}
{\bf J}=\partial_t {\bf P}=-i\omega {\bf P},
\end{equation}
one obtains in the case of dielectrics
\begin{equation}
{\bf J}({\bf r}) =- i\omega\varepsilon_0  (\varepsilon({\bf r}) - 1) {\bf E}({\bf r}),
\label{vJE}
\end{equation}
where $\varepsilon$ is the relative dielectric contrast, $\varepsilon({\bf r}) =\varepsilon({\bf r})/\varepsilon_h$. The extra dyadics is known
for a variety of different principal volumes. For instance, ${\bf L}={\bf I}/3$ for a {\em sphere}
and a {\em cube} (see Table 1 in Ref.~\cite{Yaghjian1980}). 
According to \rfs{elfvloc}-\rfs{vJE}, the measured local field is then
\begin{equation}
{\bf E}_{loc}({\bf r})={\bf E}({\bf r}) + \frac{{\bf P}}{3\varepsilon_0 }={\bf E}({\bf r})
\left(1 + \frac{\varepsilon({\bf r}) - 1}{3}\right)
=\frac{\varepsilon({\bf r})+2 }{3} \, {\bf E}({\bf r}).
\label{elfvlocm}
\end{equation}
This way the quasi-static Lorentz factor, or the Lorentz local-field correction in eq~3, is recovered.

\bibliography{references}